# Contact superconductivity in In/PbTe junctions

G. Grabecki [a,b], K. A. Kolwas [a], J. Wróbel [a], K. Kapcia [c], R. Puźniak [a], R. Jakieła [a],
M. Aleszkiewicz [a,b], T. Dietl [a,d], G. Springholz [e], and G. Bauer [e]

[a] *Institute of Physics, Polish Academy of Sciences, al. Lotnikow 32/46, PL-02-668 Warsaw, Poland*
[b] *Dept. Mathematics and Natural Sciences, College of Sciences UKSW, ul. Woycickiego 1/3 PL-01-938 Warsaw, Poland*
[c] *Faculty of Physics, Adam Mickiewicz University, ul. Umultowska 85, 61-614 Poznań, Poland*
[d] *Institute of Theoretical Physics, University of Warsaw, ul. Hoza 69, PL 00 681 Warszawa, Poland*
[e] *Institut f¨ur Halbleiter-und-Festkorperphysik, Johannes Kepler University, Altenbergerstr. 69, A-4040 Linz, Austria*

The authors report on electron transport studies on superconductor-semiconductor hybrid structures of indium and n-type lead telluride, either in the form of quantum wells or bulk crystals. In/PbTe contacts form by spontaneous alloying, which occurs already at room temperature. The alloyed phase penetrates deeply into PbTe and forms metallic contacts even in the presence of depletion layers at the semiconductor surface. Although the detailed structure of this phase is unknown, we observe that it exhibits a superconducting transition at temperatures below 10 K. Surprisingly, its resistance drops to the value expected for an ideal superconductor-normal conductor contact. Most importantly, this result indicates that the interface phase in the superconducting state becomes nearly homogeneous - in contrast to the structure expected for alloyed contacts. We suggest that the unusual interface superconductivity is linked to the unique properties of PbTe, namely, its huge static dielectric constant. Apparently the alloyed interface phase contains superconducting precipitates randomly distributed within the depletion layers, and their Coulomb charging energies are extremely small. According to the existing models of the granular superconductivity, even very weak Josephson coupling between the neighboring precipitates gives rise to the formation of a global superconducting phase which explains our observations.

## 1. Introduction

It is well known that the Andreev reflection which occurs at superconductor-normal conductor interfaces is a source of a proximity effect, which transfers superconducting correlations into the normal conductor.[1,2] If the normal part of the junction consists of a semiconductor, the induced superconductivity will be tunable by the application of gate voltage. In order to realize this idea, superconductor-semiconductor (S/Sm) structures have been studied for more than two decades.[3,4] However, the main challenge is the presence of the Schottky barrier at the interface between semiconductors and superconducting metals. The best-known exception is a narrow gap semiconductor InAs, where a naturally occurring surface accumulation layer (n-type) prevents the Schottky barrier formation.[5] For this compound, the majority of the operating S/Sm structures, like Josephson field-effect transistors[6] or electron-pair entanglers,[7] have been demonstrated. However, for further development and practical applications of these devices, a wider variety of useful S/Sm systems is desirable.

Lead telluride PbTe, a narrow gap semiconductor, can be considered as an alternative for InAs. This IV-VI compound, which is widely used in thermoelectric devices[8] and infrared detectors,[9] exhibits unusual electronic and lattice dynamical properties. It has a multi-valley band structure with nearly mirror-like conduction and valence bands edges at *L*-point of the Brillouin zone.[10] The surfaces of constant energy are four equivalent ellipsoids of revolution with their main axes oriented along the <111> directions. There is a large anisotropy of the effective masses, e.g., for the conduction band $m_\perp^* = 0.023 m_0$ and $m_\parallel^* = 0.2 m_0$. Another unique feature of PbTe is its paraelectric behavior leading to a huge static dielectric constant $\varepsilon \approx 1600$ at liquid helium temperatures.[11] This causes a screening of the Coulomb potentials of defects and, together with the small effective masses, results in very high carrier mobilities in bulk PbTe that can reach values exceeding $10^6$ cm$^2$/Vs (see Table 1). Furthermore, the effective screening reduces potential fluctuations at the nanometer scale, so that PbTe nanowires show one-dimensional conductance quantization, even if they contain quite a large number of impurities in their vicinity.[12,13] Therefore, PbTe turns out to be a good candidate for quantum ballistic devices. We expect that these unique properties might be also fruitful in the context of S/Sm junctions. In particular, one expects that the role of interface defects, which arise due to dissimilarity of the two materials, namely, the superconducting metal and the semiconductor should be significantly reduced.[5]

The goal of the present work is the experimental demonstration of these expectations using indium as a superconductor. However, in order to exploit the advantages of PbTe one has to obtain a highly transparent S/Sm junction. We have fabricated a variety of S/Sm structures using both bulk PbTe crystals as well as PbTe quantum wells (QWs). Different forms of In electrodes have been used, either thin deposited In layers or bulk In spots. They have been attached to PbTe using various methods, either by pressing (Fig. 1 a) or by employing multi-level e-beam lithography [Fig. 1(c), 1(d)]. Such a wide variety of samples allows us to draw general conclusions about the nature of phenomena taking place at the In/PbTe interface.

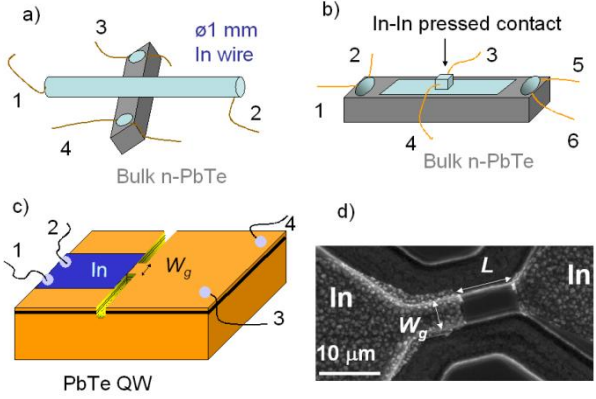

Fig. 1 Schematic representation of structures employed for investigation of the In/PbTe contact resistance: (a) pressed In contact to bulk PbTe; (b) evaporated In contact onto bulk PbTe, connected to the measuring circuit by means of a pressed In contact (c); In contact fabricated by means of electron beam lithography to PbTe/Pb$_{0.92}$Eu$_{0.08}$Te:Bi quantum well. The constriction width is $W_g$ = 30 μm. (d) SEM image of the S/Sm double junction to PbTe/Pb$_{0.92}$Eu$_{0.08}$Te:Bi heterostructure. Here $W_g$ = 6 μm and $L$ = 10 μm. A granular structure of the evaporated In layer is clearly visible.

We have found that, in contrast to InAs, a surface depletion layer is formed in n-type PbTe, which prevents a direct formation of the barrier-less contact. This was demonstrated by the deposition of pure Pb as contact. The deposition of In, however, created an excellent contact even through this barrier because it alloys with PbTe. What is most surprising, is that the alloy phase exhibits a superconducting transition at helium temperatures, which leads to a substantial reduction of the contact resistance. It approaches the limit expected for an ideally transmitting and uniform S/Sm junction. We present arguments that such behavior is a result of unique paraelectric properties of PbTe.

The present paper is organized as follows. Section 2 contains some theoretical information on S/Sm contacts and formulas used for the interpretation of the experimental data. In Section 3, existing experimental data on PbTe surface properties and In/PbTe interfaces are reviewed. In Section 4 we present properties of the starting materials and the fabrication procedures employed for obtaining the In/PbTe contacts. Section 5 contains the description of experimental findings, including structural data on the In/PbTe interfaces, results of the contact resistance measurements as a function of temperature, and the data on magnetic susceptibility. Section 6 presents a discussion of the results and their implications concerning the origin of the interface superconductivity. Section 7 contains a summary and conclusions of the paper.

## 2. Superconductor/semiconductor contacts – theory

In order to describe the conductance through metal-semiconductor contacts, in the case when the metal is either in the normal or superconducting state, one can use the theory developed by Beenakker.[14] For the normal state, the conductance is expressed by the well-known Landauer formula:

$$G_{NN} = \frac{2e^2}{h} \sum_{i=1}^{N} T_i \qquad (1)$$

where $T_i$ denotes the transmission magnitude for the $i$-th conduction channel in the contact. For the superconductor state, Eq. 1 has been modified to:

$$G_{SN} = \frac{4e^2}{h} \sum_{i=1}^{N} \frac{T_i^2}{(2-T_i)^2}. \qquad (2)$$

Here, the coefficients $T_i$ appear squared, and there is a factor 4 (instead of 2) because of the two-carrier nature of the Andreev reflection. In order to employ Eqs. 1 and 2 for an analysis of experimental data, one usually assumes that the transmission magnitudes are the same for all channels, $T_i = T$. Additionally, in the case under consideration $N \gg 1$, meaning that the quasi-classical approximation holds, so that the number of conducting modes for the three dimensional (3D) contact can be expressed as $N \approx k_F^2 A/4\pi$, where $k_F$ is the Fermi wavevector and $A$ is the contact area. Then Eq. 1 takes the form:

$$G_{NN}(3D) = 1/R_{NN}(3D) = \frac{2e^2}{h} \frac{k_F^2 A}{4\pi} T. \qquad (3)$$

For $T = 1$, this is the well-known Sharvin formula.[15] For the 2D case, $N \approx k_F W/\pi$, where $W$ is the contact width, and we have:

$$G_{NN}(2D) = 1/R_{NN}(2D) = \frac{2e^2}{h} \frac{k_F W}{\pi} T. \qquad (4)$$

If the metal is in a superconducting state, Eq. 2 is given by

$$G_{SN}(3D) = 1/R_{SN}(3D) = \frac{4e^2}{h} \frac{k_F^2 A'}{4\pi} \frac{T'^2}{(2-T')^2}, \qquad (5)$$

and

$$G_{SN}(2D) = 1/R_{SN}(2D) = \frac{4e^2}{h} \frac{k_F W'}{\pi} \frac{T'^2}{(2-T')^2}. \qquad (6)$$

The primes in Eqs. 5 and 6 indicate possible changes of the contact parameters caused by the transition to the superconducting phase. We point out that it has been assumed $A = A'$, $W = W,'$ and $T = T'$ for most of S/Sm contacts studied so far. It is easy to note from Eqs. 3-6 that if this is the case, the contact conductance $G_{NN}/G_{SN}$, increases at most by a factor of two below the superconducting transition. This occurs for $T = T' = 1$. Actually, the value of $G_{NN}/G_{SN}$ monotonically decreases when the magnitude of $T$ gets reduced becoming *smaller* than one for $T < 0.76$. This is a consequence of the fact that due to the presence of the superconducting energy gap, no one-carrier transport is possible across any S/Sm interface. Simultaneously, the two-carrier subgap transport via the Andreev reflection, which is proportional to $\sim T^2$, becomes strongly reduced when the transmission magnitude is suppressed.

This theory applies to the ballistic regime where the electron mean free path $l_e$ is much larger than the characteristic contact dimensions, $l_e >> W, \sqrt{A}$. When this condition is not fulfilled, the contact resistance also contains a contribution from diffusive transport. However, as shown by Wexler[16] and de Jong[17] for the 3D and 2D contacts respectively, the diffusion gives an additive contribution to the contact resistance in the whole range of $l_e$ values, from $l_e >> W, \sqrt{A}$ to $l_e << W, \sqrt{A}$. Since often $l_e \sim W, \sqrt{A}$, in order to extract the ballistic contribution from experimental data, the diffusive part should be independently determined and subtracted from the total contact resistance.

At an ideal S/Sm interface, $T = 1$, the efficiency of the Andreev reflection reaches 100%. Furthermore, for an ideal contact the current across the interface is rather distributed uniformly than exhibits a filamentary character, so that the effective contact width $W$ (the 2D case) or the contact area $A$ (the 3D case) are equal to those implied by the fabrication process, $W = W_g$, and $A = A_g$, respectively. We will denote the contact resistances calculated using Eqs. 3-6 for the ideal parameters as $R^*_{NN}$ and $R^*_{SN}$. Any real S/Sm structure can be, therefore, characterized by comparing the magnitude of its resistance to $R^*_{NN}$ and $R^*_{SN}$.

S/Sm structures close to the ideal limit have been only obtained for InAs and related materials.[18] This is due to the fact that there is no Schottky barrier in this case, and the interface is abrupt and spatially uniform.[5] In other S/Sm junctions where Schottky barriers are present, ohmic contacts can be obtained by alloying.[19] Typically, one (or more) metals are deposited onto the semiconductor surface and annealed in order to promote an interdiffusion process. As a result, the metal penetrates and dopes the semiconductor, and/or various interfacial reactions take place. In this way, metallic conductance is established. However, a random character of the process results in a filamentary character of the charge transport, so that usually $W << W_g$ or $A << A_g$. Consequently, the values of contact resistances are much higher than $R^*_{NN}$ and $R^*_{SN}$. The best-known example is the AuGe/Ni alloyed contact, commonly used for AlGaAs/GaAs heterostructures.[20] It has to be noted that in devices like quantum wires or quantum dots, the contact pads are placed at macroscopic distances from the active device regions. Therefore, their inhomogeneity does not play any role and a large contact pad area compensates the enhanced contact resistance. In contrast, in the case of hybrid S/Sm microdevices, the inhomogeneity is unacceptable because the contact is an essential part of the active device. This is the reason that superconducting contacts to AlGaAs/GaAs could not be obtained with a quality appropriate for useful S/Sm devices. In particular, attempts of alloying AlGaAs/GaAs with indium[21,22] or tin[23] led to rather inhomogeneous interfaces, as demonstrated by using high-resolution transmission electron microscopy (HRTEM), which revealed a number of pyramid-like filaments penetrating the AlGaAs barrier between the In layer and the 2D electron gas.[22] This nanocomposite-like structure enabled metallic conductance across the AlGaAs cap layer, however, the contact resistance was much higher than $R^*_{NN}$ and $R^*_{SN}$. In a more recent work,[24] indium contacts were prepared to GaAs/GaAlAs by employing an improved fabrication procedure. Also in this case, the contact resistance was much higher than the ideal values, and there was clear evidence that the current flew through a number of separate In filaments. This indicated that $W << W_g$. Additionally, the resistance was higher when In was superconducting than when it was in the normal state. In terms of Eqs. 4 and 6, this indicated that the transmission coefficient did not exceed 0.76. A similar behavior was also reported for many other S/Sm structures.[25]

These observations confirm that the theory of Beenakker[14] properly describes the existing S/Sm junctions provided that one takes into account the real contact width or area. It might appear that this approach can be also applied to the case of the In/PbTe contacts. However, our findings demonstrate that a dramatic increase in the values of $W$ and $A$ occurs at the superconducting transition. We link these results to unusual properties of In/PbTe interfaces, which are described in the next Section.

## 3. In/PbTe contacts: Present status

As it was mentioned before, there is a tendency in n-type PbTe to form a depletion layer at the surface. This is a result of oxidation under ambient conditions.[26] Moreover, after a sufficiently long time even p-type inversion layers are formed.[27] The depletion layer width is

$$w_{depl} = \sqrt{\frac{2\varepsilon\varepsilon_0|\Phi|}{eN}}, \qquad (7)$$

where $\Phi$ is the surface potential, $N$ the carrier concentration, and $\varepsilon$ the relative dielectric constant. Because of a huge value of $\varepsilon$ in PbTe, one expects very wide depletion layers in this material. For example, in bulk PbTe samples measured in the present work, one obtains $w_{depl}$ = 270 nm, assuming that $\Phi$ is equal to the value of the energy gap. The actual magnitude of the barrier is even higher in the case of the $Pb_{0.92}Eu_{0.08}Te$ cap layer on the PbTe QWs studied in the present paper. Therefore, obtaining of an ohmic contact to n-PbTe is possible only by an appropriate alloying.

For this purpose, we have used In. According to the available data, In forms a resonant donor in PbTe.[10] However, the electron concentration in PbTe doped with In never exceeds $10^{19}$ cm$^{-3}$. This was explained in terms of Fermi level pinning at this resonant level, which is placed about 70 meV above conduction band edge. Also important is the fact that the solubility of In in PbTe reaches 24%[28] so quite concentrated alloys can be formed. Furthermore, for In concentrations exceeding 2% the resonant level starts to shift down towards the band edge and finally enters the energy gap above 20 %. This transition is marked by an increase of the alloy resistance by several orders of magnitude.[28] For even higher In concentrations, the solid solution becomes a heterogeneous mixture of PbTe and indium tellurides, namely, InTe and $In_2Te_3$.[29] These compounds are known as wide the band gap semiconductors or insulators.[30] The above findings indicate that very strong alloying of PbTe by In leads to an insulating, rather than metallic, behavior.

What should be noted as well is the fact that the In/PbTe alloying process may occur already at room temperature. This was unambiguously shown in studies of the structure of In/PbTe contacts performed by Auger spectroscopy profiling.[31,32] For example, a significant In content was found as deep as several tenths of nm below the PbTe surface, for a contact deposited at $T$ = 80 K and heated up to 300 K for the period of 1 h only.[31] Probably, this process is responsible for the fact that In is commonly used for making low-resistance contacts to n-type PbTe.[9] However, because of the complicated behavior of the In/PbTe alloy, the contact structure is rather complex and poorly understood. In particular, the fast diffusion rate at room temperature seems to be inconsistent with the known rates for In diffusion observed during the fabrication of PbTe p-n junctions.[33] These rates were measured in the temperature range from 600°C to 1000°C, and their extrapolation to room temperature leads to negligible diffusion. Such a discrepancy indicates that the mechanisms of In diffusion at small (doping regime) and high (alloy regime) concentrations must be different. This problem was considered by Smorodina et al.,[34] who measured Auger spectra of In impurities in PbTe. They found that for small concentrations ($x << 1\%$) the bond In-Te had an octahedral character. This meant that In acted as an impurity substituting Pb. However, for $x > 1\%$, In-Te formed predominantly covalent bonds, which was characteristic for indium tellurides. Indeed, measured Auger profiles revealed the presence of these compounds in In/PbTe contacts.[31,32] According to the predictions of Sizov et al.,[35] the formation of In-tellurides is energetically favorable and might even occur at room temperature. Grishina et al.,[32] have proposed that such interfacial reactions may be like:

$$In + PbTe \rightarrow InTe + Pb + 45.68 \text{ kJ}.$$

It is strongly exothermic and can also produce Pb precipitates in the contact region. The last observation is important for the explanation of our experimental results.

Another observation, also important from the point of view of the present work, is the surprising number of superconducting phases reported for PbTe. Several authors interpreted these phases as arising from either Pb precipitation,[36,37] indium diffusion,[38] dislocation arrays [39,40] or caused by thallium (in p-type material).[41] As we discuss in Section 6, the tendency of PbTe to superconduct may be not accidental and could be a consequence of its unusual material properties. It is interesting to recall here the only existing (at least in our knowledge) study of the superconducting In layers deposited on PbTe crystals.[38] In contrast to our layers, these were deposited in ultra-high vacuum on the substrates cooled down to helium temperatures with thicknesses of the order of 10 nm. Their resistance was measured in-situ and exhibited a superconducting transition $T_c$ at about 7 K. This value is substantially higher than for the bulk indium ($T_c$ = 3.4 K). Additionally, the properties of the In layers visibly changed after even very short warming up to the room temperature. There were no structural studies, however, one could suppose that because of their small thickness the In layers were discontinuous and consisted of separate grains. Therefore, the measured resistance could contain a contribution from the In/PbTe interface region. Later on, the same authors[42] suggested that the increase of $T_c$ could be caused by an unknown superconducting phase arising due to alloying. A quite similar increase of the superconducting transition temperature has been observed in the In/PbTe contacts studied in the present work.

### 4. Experimental details

#### 4.1 Starting materials: bulk crystals and quantum wells

Bulk PbTe samples for the contact studies have been cut from a crystal grown by the Bridgman method. The electric parameters for this crystal are shown in Table 1. The *n*-type conductivity is assigned to an excess of Pb. A rather high value of electron mobility is a result of the low effective mass and the huge dielectric constant of PbTe. For contact resistance measurements, rectangular samples of dimensions 8 ×2×1.5 mm$^3$ have been prepared. Their surfaces have been mechanically polished and etched in HBr solution. The contacts are usually deposited one day after the etching process, in the meantime the samples are kept in ambient atmosphere.

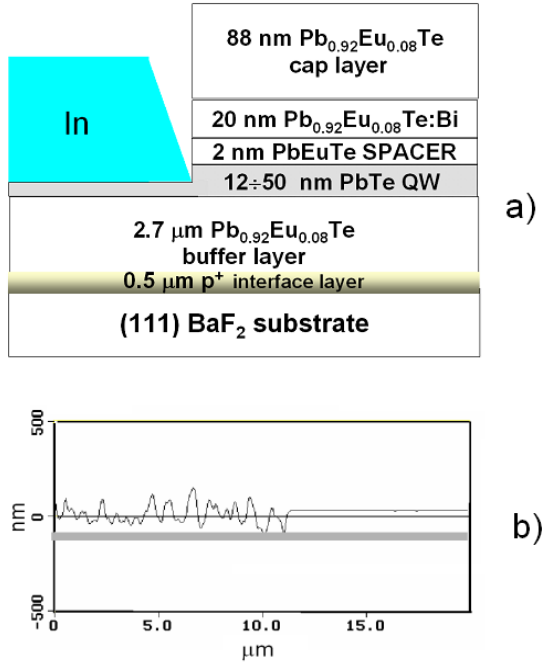

Fig. 2 (a) Layout of modulation doped PbTe/Pb$_{0.92}$Eu$_{0.08}$Te:Bi herostructure. The evaporated In layer deposited on uncovered quantum well is also shown. (b) AFM surface profile across such a structure (gray band schematically indicates the position of the PbTe quantum well).

A layout of modulation doped PbTe/Pb$_{0.92}$Eu$_{0.08}$Te:Bi heterostructures is presented schematically in Fig. 2(a). The structures have been fabricated by molecular beam epitaxy (MBE) employing protocols described previously.[43] The layers are deposited on (111) oriented BaF$_2$ substrates. A thick Pb$_{0.92}$Eu$_{0.08}$Te buffer layer separates the active PbTe QW from the heavily dislocated p-type layer that is formed at the substrate vicinity.[44] The PbTe QW resides between the Pb$_{0.92}$Eu$_{0.08}$Te barriers. For the employed Eu content of 8%, the barrier height in the conduction band is close to 180 meV.[45] In order to avoid precipitation of Eu, the MBE growth is performed with some excess of Te flux, so that the undoped QW exhibits weak p-type conductivity. Thus, to introduce electrons into the QW, a part of the front PbEuTe barrier is doped by Bi ($N_D = 3 \cdot 10^{18}$ cm$^{-3}$). The doped layer is separated from the PbTe QW by an undoped 2 nm thick Pb$_{0.92}$Eu$_{0.08}$Te spacer layer. A 90 nm thick Pb$_{0.92}$Eu$_{0.08}$Te cap layer completes the structure. In the employed layout, the n-type QW is electrically separated from the interfacial defects by the *p-n* junction. Actually, biasing of this junction can serve for changing the electron concentration in the QW. This method was previously used for tuning the electron density in quantum ballistic devices of PbTe.[12]

Basic electrical parameters of the studied QWs are shown in Table 1. As one can see, the electron mobility values are by about an order of magnitude smaller than in the bulk sample. As demonstrated in our previous work,[12] this results from alloy scattering of electrons penetrating the Pb$_{1-x}$Eu$_x$Te barriers.

### 4.2 Fabrication of In/PbTe contacts

In order to fabricate In/PbTe contacts different procedures have been employed for the bulk and QW samples. For bulk PbTe, pressed, evaporated, and soldered contacts have been studied. They are schematically represented in Fig. 1(a) and 1(b). A macroscopic In wire has been pressed towards the sample edge and fixed in a way that the thermal compression during the cooling increased the pressing force. Although the resistance of this contact exhibited different values during subsequent cooling cycles, the qualitative behavior is reproducible. The evaporated contact [Fig. 1(b)] has been defined by thermal deposition of 5N metallic In using a standard thermal evaporator. To define the contact shape, a macroscopic metallic mask is used. The vacuum during the evaporation is about 10$^{-6}$ Torr and the substrate is kept at the room temperature. The deposition time is always shorter than 40 seconds to prevent any overheating of the structure. The In layer thickness is controlled during the process by a quartz resonator and in most cases is equal to 150 nm. In order to contact the layer to the measuring circuit, a small In piece is fixed by pressing, avoiding a direct heating of the evaporated layer. In contrast, the formation of soldered contacts involve such heating up to about 150-200°C. This type of contacts is fabricated by positioning small In spots on the sample surface wetted by tiny amount of milk acid. Additionally, to determine the bulk contribution, four contacts are soldered to a rectangular sample to enable four-probe resistance measurements. It has to be noted that only for the evaporated contact, the geometrical dimensions are known accurately. For the soldered spots these can only be crudely estimated and are virtually unknown for the pressed contacts

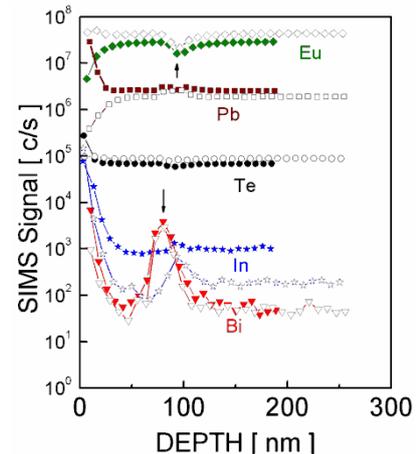

Fig. 3 SIMS analysis of the concentration profile of Eu, Pb, Te, In and Bi for one of the PbTe/Pb0.92Eu0.08Te:Bi heterostructures containing a 35 nm wide PbTe quantum well. Open symbols – data taken prior to In deposition. Closed symbols - after In deposition and subsequent dissolving of the In layer in concentrated HCl.

S/Sm structures of PbTe QWs are patterned by electron beam lithography, which allows precise definition of the contact dimension and shape. For such hybrid structures, two lithographic steps are necessary.

In the first step, insulating trenches (2 μm wide and 0.7 μm deep) are defined and etched down by means of a 0.05% Br solution in ethylene glycol. In this way, the conducting paths in PbTe QWs are defined. In the second step, the In electrodes are patterned and thermally evaporated. For this purpose, we use the same apparatus as for the contacts to the bulk PbTe. The lift-off process is employed to finally define the metal electrodes. An important point is the proper control of the vertical junction geometry because it determines the contact between the In layer and the underlying QW. We employ two possible configurations. In the first one, In is evaporated directly on the top of the cap layer, so that the superconducting electrode is separated from the QW by a 90 nm thick and 180 meV high $Pb_{0.92}Eu_{0.08}Te$ barrier. In the second one, the $Pb_{0.92}Eu_{0.08}Te$ cap layer is removed by etching and evaporated In on the uncapped PbTe QW. In this case, one expects the presence of a depletion layer due to PbTe oxidation and due to the removal of the Bi-doping layer. In Fig. 2(b), we show an atomic force microscopy surface profile of such a contact. It has to be noted that the In layers show a pronounced granular structure, characteristic for metals evaporated on uncooled substrates.[47] Because of that, their specific resistivity at room temperature is several times larger than the standard textbook value for bulk In ($8.5 \times 10^{-8}$ Ωm). Their superconducting properties have been checked in separate measurements (see next Section). It should be stressed that the samples are kept at the room temperature, and the only short-time unintentional heating could take place during the soldering process.

For our S/Sm structures we use two basic contact geometries. The first of them is a single constriction illustrated in Fig. 1(c) with the In layer placed at one side. In this case the In layer is deposited onto the uncapped QW. A scanning electron microscope (SEM) image of the other type of the structures is shown in Fig. 1(d). They consist of a PbTe wire and two In electrodes deposited on top of the cap layer. The electrodes are separated from each other by a distance $L$ of several μm. In the present work, we consider only single S/Sm interfaces, and, accordingly, $L$ is sufficiently large to exclude a Josephson coupling between them. During these studies we have fabricated about 40 structures. In Table 1 we show detailed parameters for several representative examples, for which the experimental data are discussed.

Finally, in order to study the magnetic susceptibility arising from the interface region, we have fabricated 2 × 2 millimeter arrays of 10 μm × 6 μm In rectangles deposited on top of the PbTe QWs. The distance between neighboring rectangles is 4 μm. They are evaporated into pre-etched trenches, in the same way as the structures for transport measurements. In this way, the contact area is "multiplied" to produce a measurable signal. A fragment of the array is shown as an inset to Fig. 8. Electrical parameters of the PbTe QW used for this structure are shown in lowest row of Table I.

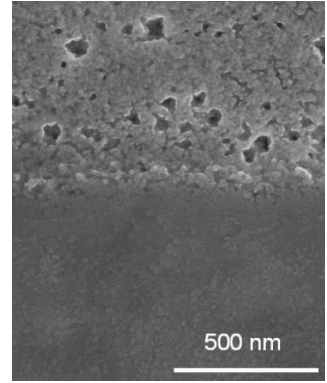

Fig. 4 Plain view SEM image of the PbTe surface after deposition of the In layer and its subsequent dissolution in concentrated HCl (upper part). The lower part shows the sample surface without In deposition but treated by HCl.

In order to determine the resistance of the In/PbTe interface, four probe method has been applied using two separate In soldered contacts attached at each side (see Fig. 1). In this way contributions from the connecting wires as well as the wire contacts are eliminated. However, for precise determination of the interface resistance, one also has to subtract the serial resistance contributions of either the In electrode or the intermediate semiconductor material. They have been independently measured by using reference structures of the same geometry. For the determination of the In layer resistance the surface of the structure is fully covered by In, and for the determination of the semiconductor contribution a sample without deposited In is employed. We have also prepared several structures where In has been substituted by Pb. Because Pb does not alloy with PbTe, this has been used to demonstrate the presence of the barrier on the surface of this material. Additionally, to check the resistance of the undoped QW, we have prepared one structure with removed both the cap and Bi-doped layers, but without In deposition. The geometry of all the reference structures is schematically presented in insets to Fig. 5 and 6.

| Sample | Starting material | $n$ [cm$^{-2}$] $l_e$ [μm] | $\mu$ [$10^4 \times \frac{cm^2}{Vs}$] | S-Sm contact type and geometry | $R_{nor}$[Ω] at 7 K | $\frac{R^*_{NN}}{R_{nor}}$ | $R_{sup}$[Ω] at 1.9 K | $\frac{R^*_{SN}}{R_{sup}}$ (T') |
|---|---|---|---|---|---|---|---|---|
| 1 | 2 | 3 | 4 | 5 | 6 | 7 | 8 | 9 |
| PbTe | n-type bulk crystal | $4.4 \times 10^{17}$ [cm$^{-3}$] | 322 | See Fig. 1(b) Evaporated In rectangle $A_g \approx 1.4$ mm$^2$ | 0.063 | $4 \times 10^{-5}$ | $4.2 \times 10^{-4}$ | 0.06 |
|  |  |  | 31 |  |  |  |  | (0.11) |
| PbTe | n-type bulk crystal | $4.4 \times 10^{17}$ [cm$^{-3}$] | 322 | See Fig. 1(b) In soldered spot $A_g \approx 0.7$ mm$^2$ | 0.025 | $2 \times 10^{-4}$ | $4 \times 10^{-5}$ | 0.13 |
|  |  |  | 31 |  |  |  |  | (0.45) |
| MBEG 1695 | PbTe/Pb$_{0.92}$Eu$_{0.08}$Te quantum well $d = 50$ nm | $7.9 \times 10^{12}$ | 26.9 | See Fig. 1(c) $W_g = 30$ μm In on PbTe QW | 221 | $4 \times 10^{-3}$ | 13 at 100 mK | 0.04 |
|  |  |  | 6.1 |  |  |  |  | (0.32) |
| MBEG 1700 | PbTe/Pb$_{0.92}$Eu$_{0.08}$Te quantum well $d = 12$ nm | $4.4 \times 10^{12}$ | 3.7 | See Fig. 1(d) In on cap $W_g = 6$ μm $L = 10$ μm | 63.1 | 0.21 | 18.6 | 0.35 |
|  |  |  | 1.3 |  |  |  |  | (0.74) |
| MBEG 1700 | PbTe/Pb$_{0.92}$Eu$_{0.08}$Te quantum well $d = 12$ nm | $4.4 \times 10^{12}$ | 3.7 | See Fig. 1d In on cap $W_g = 6$ μm $L = 3$ μm | 42.6 | 0.31 | 17.9 | 0.36 |
|  |  |  | 1.3 |  |  |  |  | (0.75) |
| MBEG 481 | PbTe/Pb$_{0.92}$Eu$_{0.08}$Te quantum well $d = 12$ nm | $6 \times 10^{11}$ | 7.5 | See Fig. 1c $W_g = 30$ μm Pb on PbTe QW | >10$^7$ |  | >10$^7$ |  |
|  |  |  | 1.0 |  |  |  |  |  |
| MBEG 1703 | PbTe/Pb$_{0.92}$Eu$_{0.08}$Te quantum well $d = 35$ nm | $3.6 \times 10^{12}$ | 17.2 | Array of In rectangles on PbTe QW See Fig. 8 inset |  |  |  |  |
|  |  |  | 2.7 |  |  |  |  |  |

Table 1. Summary of In/PbTe hybrid structure parameters and the interface resistance drop between $T = 7$ K and $T = 1.9$ K. In the last columns, the values of the transmission coefficient in the superconducting state $T'$ determined from the experimental data are shown in brackets. Notations: $n$ electron concentration, μ - mobility ($l_e$ electron mean free path), $W_g$ – 2D apparent contact width, $A_g$ - 3D apparent contact area, $R_{nor}$ – measured interface resistance in the normal state and $R_{sup}$ – measured interface resistance in the superconducting state. $R^*_{NN}$ and $R^*_{SN}$ are interface resistances for ideal S/Sm contacts, in normal and superconducting state, calculated from Eqs. 3-6.

The contact resistance measurements as a function of temperature have been performed using either a He$^4$-vapor cooled variable temperature insert or a plastic He$^3$/He$^4$ dilution refrigerator, both immersed in a standard helium cryostat equipped with a superconducting coil capable of producing fields up to 9 T. Lock-in technique with frequency 18.6 Hz is employed for the resistance measurements. In the present work, only the linear transport regime is studied. For this reason, the sample currents are kept in the nanoampere range and the measured voltages are smaller than 50 μV. The measurements of ac-magnetic susceptibility are performed in the temperature range 2 - 10 K by using the in-plane magnetic field of the amplitude of 0.1 - 10Oe and frequency of 100 Hz - 10 kHz We have found that the obtained results are independent of the magnitude and frequency of the ac field in this range.

## 5. Experimental Results

### 5.1 Structural properties of In/PbTe contacts

In order to confirm that an alloying process occurred in our structures, we have examined the compositional profile by means of secondary ion mass spectroscopy (SIMS). Oxygen ions have been employed as the primary beam. Primary ions energy is set to 12.5 keV with the beam current keep at 60nA. The secondary ions are collected from a central region of 60 μm in a diameter. Figure 3 shows a representative result obtained for the PbTe/Pb$_{0.92}$Eu$_{0.08}$Te layer no. 1703. This is the same layer for which magnetic susceptibility has been studied, see Table 1. The data reflect the multilayer sequence [see Fig. 2(a)]. In particular, one can recognize the position of the Bi-doped layer and of the PbTe quantum well (indicated by arrows). Unexpectedly, the initial layers also contain a not negligible amount of indium. Most probably, this is a result of unintentional incorporation in the MBE chamber, where the BaF$_2$ substrate is fixed using this metal to the substrate heater.[48] For this reason, we have performed additional tests to check whether or not this initial amount of indium causes any unintentional superconductivity.[49] However, neither the magnetic susceptibility (see Fig. 8) nor the transport data showed any traces of superconductivity in the initial layers. It is important to note here that the bulk PbTe crystals does not contain any initial In content because they are grown using a completely different method.

A SIMS analysis of the sample with the deposited In layer represented by full symbols in Fig. 3. In this case, the In layer has been dissolved in concentrated HCl prior to the analysis. Otherwise, the temperature increase caused by the incident ion beam melts indium and floods the analyzed area. Since immersing in HCl has only minor effect on the surface morphology and the electrical properties of the PbTe/Pb$_{0.92}$Eu$_{0.08}$Te layers, one may expect that such procedure does not influence the contact structure. As one can see from Fig. 3, there is only a moderate increase of the In concentration after this process. However, a more pronounced effect is the increase of the lead concentration near the surface. This may be a result of the interfacial reactions considered in the Sec. 3.

## 5.2. Transport measurements

Before presentation of the main results on the contact resistance in the S/Sm structures in the superconducting regime, we show several results of transport measurements performed on reference samples. They show contributions from different constituents of our structures to the measured junction resistance and prove the existence of a surface barrier. These results allow us to assign unambiguously the main effect to the In/PbTe interface. Figure 5 shows the resistance of the continuous In layer as a function of temperature (curve 1 - lower axis) and magnetic field (curve 2 -upper axis), for the reference constriction of the geometry as seen in Fig. 1(c), but fully covered by a 150 nm thick In layer (see Inset). Apparently, the superconducting transitions are quite abrupt. There is only a minor broadening and the superconducting transition temperature of $T_c \approx 3.9$ K, is several tenths of Kelvin higher than the textbook value for indium (3.4 K). Also, the critical magnetic field is slightly reduced with respect to the bulk In value (29 mT) and only about 90% of the normal resistance is restored up to magnetic fields of 100 mT. We attribute these anomalies to result from the granularity of the In layer. We have also checked that the superconducting state is preserved at least for the currents of 1 mA passing through the structure. This value is much larger than all currents used for the measurements in this work.

In Fig. 6 the resistance of several reference structures of the geometry corresponding to that shown in Fig. 1 c, but with different modifications indicated by schemes placed at the curves. All data are shown as a function of temperature in the range from 77 to 300 K, in the form of an Arrhenius plot. Curve 1 represents the data for an empty constriction, patterned in the PbTe/Pb$_{0.92}$Eu$_{0.08}$Te QW. It shows metallic behavior in the whole temperature range. The maximum near 200 K is a result of the buffer layer conduction, which short-circuits the insulating trenches at higher temperatures.[50] Curve 2 represents the resistance of the In contact to uncovered PbTe QW. Although the resistance is increased with respect to the empty structure, it also shows metallic behavior. Curves 1 and 2 are almost parallel to each other and their difference is only weakly temperature dependent. Curve 3 represents the resistance of the structure with removed cap and Bi-doped layer on one side of the constriction. The resistance exhibits insulating behavior at low temperatures. In particular, its increase is approximately exponential, with an activation energy $E_a \approx 100$ meV, which is nearly half of the energy gap in PbTe. Finally, curve 4 shows the resistance of a structure where In has been replaced by Pb. Quite surprisingly, there is also insulating behavior visible, and the resistance increases almost in parallel to that observed in the previous structure. The above results confirm the existence of a depletion layer at the uncovered surface of the PbTe QW. This barrier is removed by deposition of In, because this metal is able to alloy with PbTe, whereas alloying does

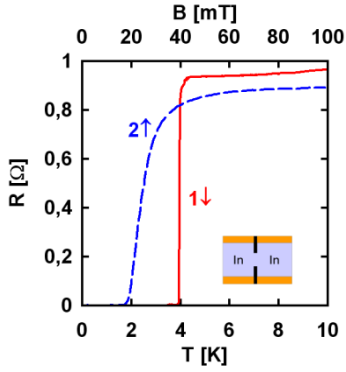

Fig. 5. Resistance as a function of temperature (curve 1 - lower scale) and magnetic field (curve 2 - upper scale) for the 30 μm wide constriction patterned in a PbTe/Pb0.92Eu0.08Te:Bi heterostructure and fully covered by a 150 nm thick In layer. This is the reference structure for the one presented in Fig. 1(c).

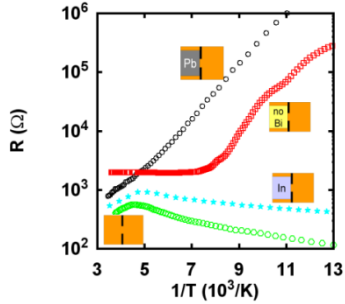

Fig. 6. Resistance as a function of the inverse temperature for the S/Sm structure presented in Fig. 1(c), compared to results obtained for several reference structures. Curve 1: constriction patterned of the PbTe/Pb$_{0.92}$Eu$_{0.08}$Te:Bi heterostructure. Curve 2: In/PbTe junction containing the In layer deposited on top of the uncovered PbTe quantum well on one side of the contact [see no.1695 in Table I and Fig. 2(b)]. Curve 3: empty constriction with uncovered PbTe quantum well on one side and without the In layer. Curve 4: Pb layer (instead of In) deposited on top of the PbTe quantum well (MBEG 481 in Table l).

It has to be noted that because the SIMS analysis averages the signal over an area with a diameter of about 60 μm, one cannot resolve whether In and Pb is distributed uniformly or whether these metals form tiny inclusions. We have performed investigations of the sample surface after dissolution of the In layer in HCl, using field emission scanning electron microscope. An example image obtained for an array of In rectangles (sample 1703 – see Table 1) is represented in Fig. 4. One notices the distinct difference between the unperturbed PbTe surface and the one previously covered by In. While the former is very smooth, the latter shows a porous structure. There are numerous holes of dimensions up to several tenths nm, indicating that HCl etching rate has been very fast in these regions. The holes are surrounded by the regions much more resistive to the acid. Such picture indicates a chemical disorder of the surface, which implies inhomogeneity of the electrical properties of the In/PbTe interface. This observation is in agreement with the established picture for alloyed contacts.

not take place in the case of Pb. This is also confirmed for samples with Pb deposited on the cap layer of PbTe/Pb$_{0.92}$Eu$_{0.08}$Te QWs.

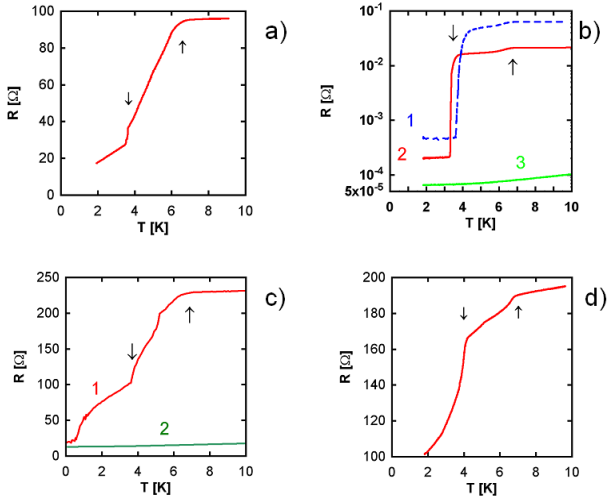

Fig. 7. Low temperature resistance for various In/PbTe structures. Plots denoted by a-d, corresponds to the structures shown in Figs. 1(a-d). The symbol $R_{ij,lk}$ indicates that the current and voltage probes are $i, j$ and $l,k$, respectively. (a) $R_{14,32}(T)$ for In contact pressed to the bulk PbTe. (b) Curve 1: $R_{34,56}(T)$ measured between the evaporated In contact and one of the soldered contacts, Curve 2: $R_{12,56}(T)$ measured between two soldered contacts. Curve 3 represents the resistance of the bulk PbTe. (c) Curve 1: $R_{13,24}(T)$ for PbTe/Pb$_{0.92}$Eu$_{0.08}$Te:Bi heterostructure with an In layer deposited on the uncovered quantum well (MBEG 1695 in Table 1). Curve 2 is for the reference structure without In. (d) $R(T)$ for PbTe/Pb$_{0.92}$Eu$_{0.08}$Te:Bi heterostructure with two In layers deposited on the top of PbEuTe cap layer (MBEG 1700 in Table I). Vertical arrows indicate steps associated with interfacial superconducting transitions.

Figure 7 represents the main result of this work, namely, the resistance for several S/Sm structures as a function of temperature in the range below 10 K. Four plots marked by letters a-d, correspond to the four different structures shown in Fig. 1, marked by corresponding letters. For all cases, we observe qualitatively the same behavior: An abrupt resistance drop with decreasing temperature occurring in form of two characteristic steps. The first step starts to appear in the temperature range between 6 and 7 K, while the second one becomes apparent between 3 and 4 K. The latter corresponds quite well to the superconducting transition in the In layer. Importantly, the magnitude of the step is much larger than the resistance of the In electrode (see Fig. 5) in its normal state, indicating that this result cannot simply be explained as originating from the vanishing of the In resistance below $T_c$.

Because the observed resistance drops are very dissimilar to those occurring in pure In, we conclude that they have to be assigned to the interface region between In and PbTe. The main effect of this "interface superconductivity" is a substantial reduction of the contact resistance. We describe it quantitatively in terms of Eq. 3-6. To employ the theoretical expressions we need to extract the interface resistance from the measured values that contain contributions from the series resistance of contact pads and the diffusive transport in the intermediate semiconductor material. The best way to proceed is to use the reference data for empty structures. We illustrate the procedure for the single S/Sm constriction whose resistance is shown in Fig. 7(c). As one can see, its resistance drops from $R_1$ = 229 Ω at $T$ = 7 K to $R_2$ = 19 Ω at 100 mK (curve 1). Simultaneously, the resistance of the reference constriction without In monotonically decreases from 16 to 13 Ω (curve 2). It contains two contributions (in series) of the large connecting pads and the constriction resistance. As was mentioned in Section 2, the latter may be approximated by a sum of diffusive and the ballistic parts. For a constriction of the width $W_g$ = 30 μm, and the parameters of the layer (MBEG 1695), Eq. 4 gives a ballistic contribution $R^*_{NN}$ = 0.96 Ω. Then the contact pad and diffusive resistances sum up to a value of about 12 Ω. Because for the S/Sm structure half of the constriction is covered by the In layer of negligible resistance, for the second half we adopt the value of 6 Ω. When one subtracts this value from the measured magnitudes of $R_1$ and $R_2$, the interface resistances are $R_{nor}$ =221 Ω and $R_{sup}$ =13 Ω for the normal and the superconducting states of the In electrode, respectively. The determination of $R_{nor}$ and $R_{sup}$ for the structure with double S/Sm junction [Fig. 1(d)] is even more simple. Because the contact pads in this case are covered by an In layer and their contribution to resistance is negligible, one has only to subtract the diffusive resistance of the Sm wire placed between the two S/Sm interfaces. Less accurate is determination of $R_{nor}$ and $R_{sup}$ for In contacts to bulk PbTe because their geometry is less precisely defined. In this case, we have subtracted the bulk contribution, which has been independently determined by the four-probe method. The values of $R_{nor}$ measured at 7 K, and $R_{sup}$ measured at 1.8 K are listed in in Table 1. It has to be noted that a quantitative analysis is not possible for the pressed contact because of its unknown dimensions.

In order to compare the results to the ideal contact resistances $R^*_{NN}$ and $R^*_{SN}$, the calculated ratios $R^*_{NN}/R_{nor}$ and $R^*_{SN}/R_{sup}$ are listed in Table 1. It has to be remembered that $R^*_{SN}$ already contains the factor of two due to the Andreev reflection. Then for the ideal contact one expects $R^*_{NN}/R_{nor}$ = $R^*_{SN}/R_{sup}$ = 1. However, for a non-ideal contact, one has always $R^*_{NN}/R_{nor}$ > $R^*_{SN}/R_{sup}$. The fact that we observe, as listed in Tab. 1, the opposite inequality indicates that the resistance values are modified by the interface superconductivity. In terms of Eqs. 3-6, we may assume that either the real contact dimensions ($W$ in 2D case or $A$ in the 3D case) or the transmission coefficients $T$ are enhanced. Let us consider this conjecture, for example, for the 2D contacts. There are two limiting cases:

1) If one assumes $T$ = $T'$ =1, then the numbers in columns 7$^{th}$ and 9$^{th}$ represent relative contact widths $W/W_G = R^*_{NN}/R_{nor}$ and $W'/W_G = R^*_{SN}/R_{sup}$. The resistance drop at the superconducting transition is exclusively caused by an increase from $W$ to $W'$. Therefore, $W$ and $W'$ are lower bounds of the real contact widths in the normal and superconducting state, respectively. 2) If one assumes that $W = W' = W_g$, the resistance drop is caused exclusively by an increase of the transmission from $T$ to $T'$. For the normal state one now has $T = R^*_{NN}/R_{nor}$, however, for the superconducting state, Eq. 6 yields $T' = 2q/(1+q)$, where $q = \sqrt{R^*_{SN}/R_{sup}}$. The values of $T'$ determined under assumption 2) are shown in brackets in column 9. In this case, $T$ and $T'$ are lower bounds for the transmissions of the S/Sm contacts. The same reasoning holds for 3D-contacts, the only change is substituting $W$ and $W'$ by the effective contact areas, $A$ and $A'$.

Characteristically, the values of $R^*_{NN}/R_{nor}$ for the contacts fabricated of the bulk PbTe and to the uncapped QWs (no. 1695) are much smaller than those with In deposited on the $Pb_{0.92}Eu_{0.08}Te$ cap layer. The latter is two orders of magnitude larger than the former. This confirms the fact that the depletion layer which exists on the bare PbTe surface produces a wider barrier. However, in the superconducting state, the respective differences between the values of $R^*_{SN}/R_{sup}$ become significantly smaller. Furthermore, the transmissions $T'$ are similar for all structures.

It has to be stressed that despite the calculated values of $T'$ are the lower bounds for the transmission in the superconducting state, they are already in the range from 0.1 to 1. This means that they compete with the best transmission values reported for other S/Sm systems.[5] Furthermore, if one assumes that $W' < W_g$ ($A' < A_g$), they must be even higher. At the lowest possible $W'$ (or $A'$) set by the upper numbers in 9th column of Tab. 1, $W'/W_G = R^*_{SN}/R_{sup}$, one has to assume $T' = 1$. Therefore, the superconducting transition occurring in the interface region would produce contacts of an exceptional quality.

5.3. Magnetic susceptibility measurements

One important confirmation of the presence of an interface superconducting phase in our S/Sm structures is due to magnetic susceptibility measurements. Such data have been obtained for a large array of In rectangles deposited on uncovered PbTe QW (for the initial layer parameters, see lowest row in Table 1). Figure 8 represents the results obtained using a parallel ac magnetic field, for the sample with indium rectangles and for an empty reference QW. In both cases we observe a continuous decrease of the susceptibility with increasing temperature, proportional to $1/T$. It stems from the paramagnetic contribution of the Eu ions present in the $Pb_{0.92}Eu_{0.08}Te$ barrier, cap and buffer layers. However, two additional jumps arise on this background for the layer with In rectangles as indicated by arrows in Fig. 8. These are interpreted as diamagnetic contributions from superconducting phases formed at the In/PbTe interface. Indeed, they appear at 6 K and 3.5 K, which agree quite well with the steps in the transport data. There is also a further, much smaller step at about 7.5 K. On the contrary, no anomalies are observed for the empty QW, i.e., not covered by In, which confirms that it does not contain any superconducting phases.

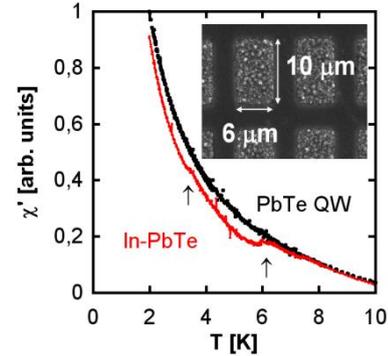

Fig. 8. Real part of ac magnetic susceptibility measured on a large array of 6×10 μm2 In rectangles deposited on uncovered, 35 nm wide PbTe quantum well in PbTe/Pb0.92Eu0.08Te:Bi heterostructures (MBEG 1703, see Table I). The susceptibility of the unprocessed heterostructure is shown for comparison. Inset: SEM image of the In array.

One has to notice that the signal from the indium layer (with the onset expected at about $T_c$ = 3.4 K) is quite weak, probably because the layer is thinner than the superconductor penetration length λ. We have also checked that in the perpendicular ac field configuration the diamagnetic contribution from indium becomes very pronounced, by far exceeding that of the interface superconducting phases. This is an additional evidence that the In layers preserve their superconducting properties after being patterned into micrometer size patches.

6. **Discussion**

The main observations on In/PbTe contacts are as follow. Apparently a large barrier exists at the uncovered n-type PbTe surface. According to literature data,[26,27] this results from surface oxidation, which causes the carrier depletion. The barrier height is actually greater than that formed by the larger-energy gap $Pb_{0.92}Eu_{0.08}Te$ surrounding the QWs whose parameters are known: its height is 150 meV and the thickness is 90 nm. A deposition of metallic Pb leaves the interfacial barrier intact. The situation is completely different for metallic In which diffuses into PbTe already at room temperature. Thus, an ohmic contact is formed owing to the In penetration into the barriers, either associated with the $Pb_{0.92}Eu_{0.08}Te$ alloy or the surface oxide.

The SIMS data suggest that, apart from the diffusion of In, the alloying process causes precipitation of pure lead or lead rich compounds. Microscopic observations of the PbTe surface after dissolving the In layer, suggest a strongly inhomogeneous contact morphology. Accordingly, the results of transport measurements at $T$ > 7 K show that the magnitude of the contact resistance is significantly larger than that expected in the ideal limit, $R^*_{NN}$. This indicates that $T << 1$ and/or $W << W_g$ (in the 2D case), or $A << A_g$ (in the 3D case). Such features are well known for alloyed contacts in other semiconductor systems (see Sec. 2).

The unexpected result is the strong resistance drop in the temperature range below 10 K. We interpret this effect in terms of a superconducting transition occurring in the interfacial region. In this way, the initial S/Sm structure changes into S/S'/Sm type, *i.e.*, consists of two serially connected superconducting regions: the In layer - S, and the interface phase - S'. The transition shows an onset near $T = 7$ K, which is significantly above $T_c$ of indium. Possibly, this may be connected to the presence of additional lead at the interface, as suggested by the SIMS data. The most important consequence of the formation of the phase S' is the reduction of the contact resistance to values close to those expected for an ideal Andreev contact, $R_{SN}^*$. This indicates a substantial increase of the contact transmission and an increase of its real width (area). One has to note that even if we assume that the contact width (area) is equal to that defined by the fabrication process, the transmission coefficients remain comparable to those observed in the best S/Sm junctions studied up to now. This strongly suggests that the S' phase is homogeneous, in a stark contrast with the known picture of the alloyed contacts of other semiconductors, as discussed in Sec. 2.

The question arises why in the case of In/PbTe a rather homogeneous superconducting phase S' is formed in a strongly disordered region. We answer this question considering a possible interface structure and the unique properties of PbTe. Most probably, the interface region can be treated as consisting of a percolating array of In and/or Pb precipitates, randomly distributed within the depleted region of PbTe. It is also possible that some of percolating paths are related to structural defects like dislocations, which may enhance the diffusion of In. For example, it was established that dislocations enhanced the diffusion rate in the Ti/Al/Mo/Au alloyed contacts to AlGaN/GaN heterostructures.[51]

Regardless of a microscopic geometry, such arrays resemble granular electronic systems,[52,53] particularly those consisting of superconducting grains distributed in a non-superconducting medium. The theories reviewed in Ref. 53 predict that a global superconductivity in such a system is possible due to Cooper pair tunneling between neighboring grains. Furthermore, it was found that the tunneling is effective only if the Coulomb charging energies of the grains are small comparing to the Josephson coupling energies,[54] the role of the Coulomb interactions being particularly important if the precipitates are placed in an insulating matrix.[55]

Because of the huge dielectric constant of PbTe at low temperatures ($\varepsilon > 1000$), the Coulomb charging energy of such precipitates is expected to be about two orders of magnitude smaller than in other systems such as GaAs. The reduction of the charging energy enables the development of a global superconductivity for much weaker Josephson couplings.

# 7. Conclusions

We have studied the electron transport in superconductor-semiconductor hybrid structures of indium and n-type PbTe, either in the bulk or quantum well form. In this system, metallic contacts are formed by spontaneous alloying already at room temperature. The metallic phase penetrates deeply into the semiconductor volume, which allows for the formation of the ohmic contacts even in the presence of a large barrier at the interface. Although the detailed chemical and crystalline structure of the resulting interface alloyed phase remains unknown, we have found that it exhibits a broad range of superconducting transitions, occurring in the temperature region between 4 and 7 K. The appearance of superconductivity at temperatures above Tc of pure In is independent of the junction fabrication procedure and causes a substantial drop in the contact resistance below 10 K. Actually, the value of the contact resistance becomes comparable to that predicted for an ideal and spatially uniform superconductor-normal junction. This result indicates that the superconducting interface phase is quite homogeneous - in contrast to expectations for alloyed contacts specified by filamentary conductance in the normal state.

We put forward a conjecture that the uniform interface superconductivity is a result of the paraelectric character of PbTe at helium temperatures. Owing to the huge magnitude of the dielectric constant that reduces significantly the Coulomb charging energy, the global superconducting phase spreads over the whole interface region even for rather small magnitudes of the Josephson couplings between the superconducting grains involving In and/or Pb. One practical result of our work is that a rather transparent S'/Sm interface is obtained with the alloyed contacts, despite of their necessarily inhomogeneous structure.

Possibly, such a mechanism, together with high values of electron mobility, may explain the appearance of superconductivity in PbTe, as reported by many authors.[36-41] This is an interesting possibility because it potentially allows tuning the superconducting properties within a semiconductor, *e.g.*, by changing a distance between the precipitates. A similar bulk-like superconductivity induced by precipitates of Hg (or perhaps FeSe) was observed in (Hg,Fe)Se.[56].

Acknowledgements

The authors are indebted to Tomasz Story for providing a high-quality n-type PbTe bulk crystal. This work was supported by projects: Ministry of Science and Higher Education (Poland) grant no. 1247/B/H03/2008/35, ESF Project no. 107/ESF/2006 (SPINTRA), and the Austrian Science Funds (ISO-N20).